# PC-Cluster based Storage System Architecture for Cloud Storage

Tin Tin Yee[1] and Thinn Thu Naing[2]

[1]University of Computer Studies, Yangon, Myanmar
`tintinyee.tty@gmail.com`
[2]University of Computer Studies, Yangon, Myanmar
`thinnthu@gmail.com`

## ABSTRACT

*Design and architecture of cloud storage system plays a vital role in cloud computing infrastructure in order to improve the storage capacity as well as cost effectiveness. Usually cloud storage system provides users to efficient storage space with elasticity feature. One of the challenges of cloud storage system is difficult to balance the providing huge elastic capacity of storage and investment of expensive cost for it.*

*In order to solve this issue in the cloud storage infrastructure, low cost PC cluster based storage server is configured to be activated for large amount of data to provide cloud users. Moreover, one of the contributions of this system is proposed an analytical model using M/M/1 queuing network model, which is modelled on intended architecture to provide better response time, utilization of storage as well as pending time when the system is running. According to the analytical result on experimental testing, the storage can be utilized more than 90% of storage space. In this paper, two parts have been described such as (i) design and architecture of PC cluster based cloud storage system. On this system, related to cloud applications, services configurations are explained in detailed. (ii) Analytical model has been enhanced to be increased the storage utilization on the target architecture.*

## 1. INTRODUCTION

Cloud Computing is an emerging computing platform and service mode, which organize and schedule service based on the Internet. In generally, the concept of cloud computing incorporate web infrastructure, Web 2.0, virtualization technologies and other emerging technologies. With the cloud computing technology, users use a variety of devices, including PCs, laptops, smart phones, and PDAs to access programs, storage, and application-services offered by cloud computing providers. Advantages of the cloud computing technology include cost savings, high availability, and easy scalability.

The main objective of cloud computing is to provide ICT services over the cloud. The service models are divided in Cloud as Software-as-a-Service (SaaS), which allows users to run applications remotely from the cloud. Infrastructure-as-a-service (IaaS) refers to computing resources as a service. This includes virtualized computers with guaranteed processing power and reserved bandwidth for storage and Internet access. Platform-as-a-Service (PaaS) is similar to IaaS, but also includes operating systems and required services for a particular application. In other words, PaaS is IaaS with a custom software stack for the given application. The Data storage-as-a-Service (DaaS) provides storage that the consumer is used including bandwidth requirements for the storage.

From the end user point of view, cloud computing services provide the application software and operating system from the desktops to the cloud side, which makes users be able to use anytime from anywhere and utilize large scale storage and computing resources. In Cloud Computing, disk storage is the one of the biggest expenses. There is strong concern that cloud service providers will drown in the expense of storing data, especially unstructured data such as documents, presentations, PDFs, VM Images, Multimedia data, etc. Cloud service providers offer huge capacity cost reductions, the elimination of labour required for storage management and maintenance and immediate provisioning of capacity at a very low cost per terabyte. Therefore, the storage and computing on massive data are major key challenge for a cloud computing infrastructure. In this paper, we focus on PC cluster based storage server which provides the hardware and software devices for large amount of data storage.

The rest of this paper is organized as follows. Section 2 describes cloud storage and related works. In section 3 presents Academic based Private Cloud architecture. In section 4 discuss how to build physical topology of CCPS (Cloud based Cluster PC System). In section 5 describes analytical model and experimental environment presented in section 6. Finally, section 7 is the conclusion and future work.

## 2. CLOUD STORAGE AND RELATED WORKS

The Cloud has become a new vehicle for delivering resources such as computing and storage to customers on demand. Cloud storage is one of the services which provide storage resource and service based on the remote servers based on cloud computing. Cloud storage will be able to provide storage service at a lower cost and more reliability and security. The advantage of cloud storage is it enables users at any time access data.

Nowadays, a huge variety of cloud storage systems is available, all with different functionality, optimizations and guarantees. That is, solutions focus on a particular scenario and the provided services are tailored accordingly. As a result, cloud storage systems vary in the data format (e.g., Key/Value vs. row store), access-path optimization (e.g., read vs. write, one-dimensional vs. multi-dimensional access), distribution (e.g., single vs. multi-data center distribution), query language, transaction support, availability. Typical cloud storage system architecture includes a master control server and several storage servers. Many cloud storage architecture focus on multiple tenants.

Hussam [4] described RAID (Redundant Arrays of Inexpensive Disks)-like techniques at the cloud storage and showed that reduce the cost and better fault tolerant. Amazon S3 provides a simple web services interface that can be used to store and retrieve any amount of data, at any time, from anywhere on the cloud. It gives any developer access to the same highly scalable, reliable, secure, fast, inexpensive infrastructure that Amazon uses to run its own global network of web sites. The service aims to maximize benefits of scale and to pass those benefits on to developers [1]. André [5] presents analytical considerations on the scalability of storage clusters and presents a storage cluster architecture based on peer-to-peer computing that is able to scale up to hundreds of servers and clients. It used Internet SCSI (iSCSI) as inter-connect protocol. This storage cluster environment is implemented and tested on a Linux based HPC-cluster. Yong [6] proposed a storage cluster architecture CoStore using network attached storage devices. In CoStore the consistency of a unified file namespace is collaboratively maintained by all participating cluster members without any central file manager and designed the data anywhere and metadata at fixed locations file system layout for efficiency and scalability in a CoStore cluster. Jiyi [9] introduced cloud storage reference model. This model designed scalable and easy to manage storage system but aren't designed to be high performance. This paper presented the key technologies, several different types of clouds services, the advantages and challenges of cloud storage. Xu [10] proposed new cloud storage architecture based on P2P which provide a pure distributed data storage environment without any central entity. The cloud based on the proposed architecture is self-organized and self-managed and as better scalability and fault tolerance using Distributed Hash Table approach.

## 3. ACADEMIC BASED PRIVATE CLOUD ARCHITECTURE

The main focus of this system is to configure a cloud storage system using PC cluster in order to utilize on academic based private cloud. All cloud configurations are implements with Ubuntu Enterprise Cloud (UEC) architecture. The UED is powered by Eucalyptus [2], an open source implementation for the emerging standard of the Amazon EC2 API. The intended architecture is shown in figure 1.

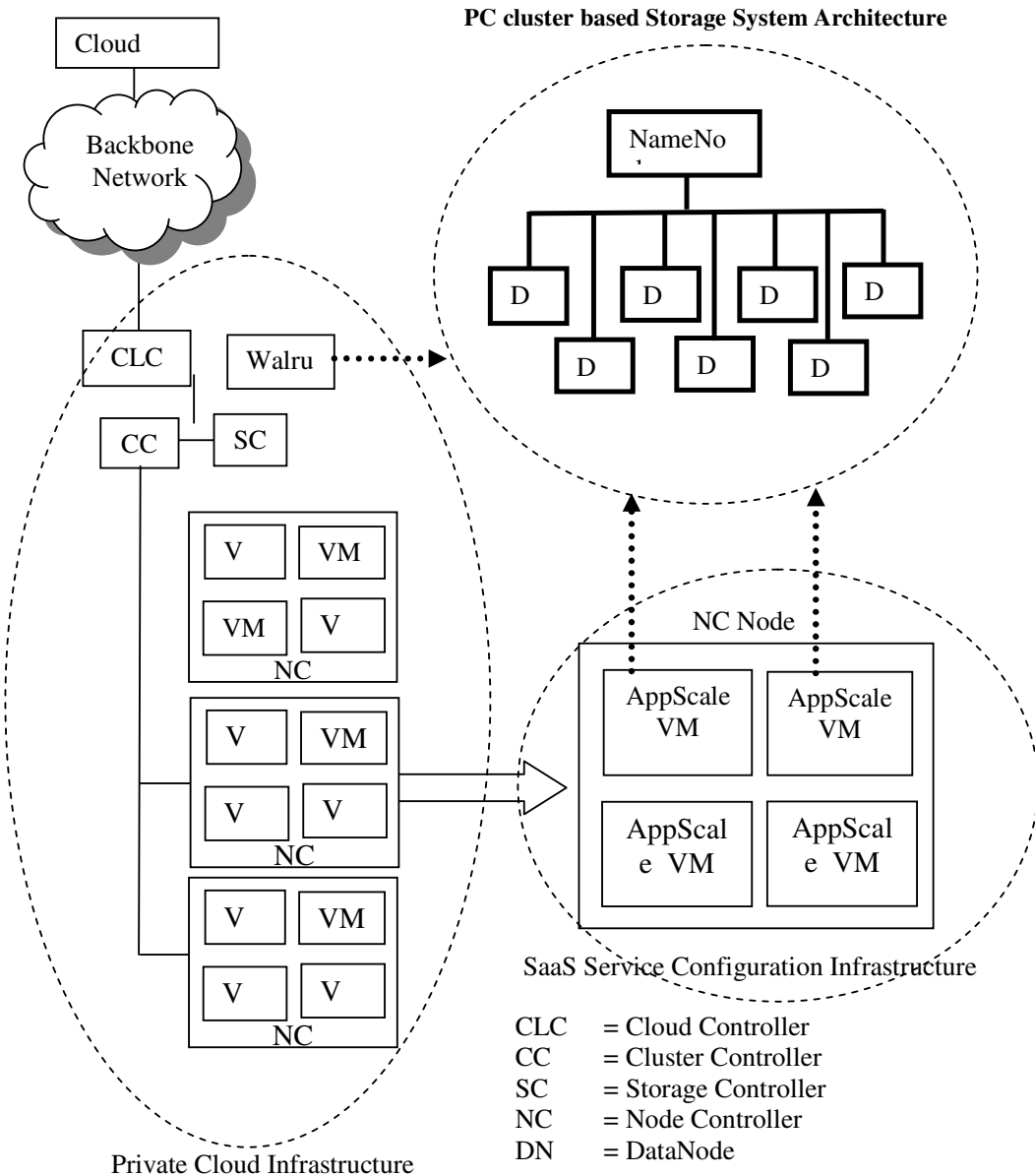

Figure 1. Architecture of Academic based Private Cloud

Physical network architecture is based on the backbone network. This architecture consists of front-end infrastructure, back-end infrastructure and PC cluster based storage server. The front-end infrastructure contains Cloud Controller (CLC), Cluster Controllers (CC), and Walrus. The

119

back-end infrastructure consists of Node Controllers (NC). The PC cluster based storage server is makes up of many of PC cluster node that is the focus of our work. Cloud testbed systems configuration is shown in table 1 and table 2 present allocated IP address.

### 3.1. Front-End Infrastructure

The front-end infrastructure contains Cloud Controller (CLC), Cluster Controllers (CC) and Walrus Storage Controller (WSC). And then, Elastic Block Storage Controller (EBS) runs on the same machines as the Cluster Controller and is configure automatically when the Cluster Controller. It allows to create persistent block devices that can be mounted on running machines in order to gain access to virtual hard drive.

The Cloud Controller (CLC) is the entry-point into the cloud for users and administrators. It queries node managers for information about resources, makes high-level scheduling decisions, and implements them by making requests to cluster controller. The Cluster Controller (CC) operates as the go between the Node Controller and the Cloud Controller. It will receive requests to allocate virtual machine images from the Cloud Controller and in turn decides which Node Controller will run the virtual machine instance as well as manages virtual instance network. All the virtual machine images were stored in the Walrus that providing a mechanism for storing and accessing virtual machine image.

### 3.2. Back-End Infrastructure

Back-end infrastructure consists of Node Controllers (NC). A Node Controller executes on every node that is designed for hosting AppScale VM instances. These AppScale VM instances has own IP address on Node Controller. Each of Node Controller has a local storage device. The local storage was only used to hold virtual machine image at run time and for caching virtual machine instances. When an AppScale VM instance is terminated, Eucalyptus will not save the data inside the AppScale VM instance, it needs an external storage server to save these application data in the AppScale VM instance. Therefore, before terminating an instance, user must upload the data to storage server.

Table 1. Cloud testbed system configuration

| System | Process Configuration | Other Information |
|---|---|---|
| Cloud Controller (CLC) | Intel ® Core™ 2 Duo | Processor: Intel® Core ™ 2 Duo CPU E7500 @ 2.93GHz<br>Memory:2GB RAM<br>Storage: 250GB HDD |
| Cluster Controller (CC) | Intel ® Core™ 2 Duo | Processor: Intel® Core ™ 2 Duo CPU E7500 @ 2.93GHz<br>Memory:2GB RAM<br>Storage: 250GB HDD |
| Node Controller (NC) | Intel ® Core™ 2 Duo | Processor: Intel® Core ™ 2 Duo CPU E7500 @ 2.93GHz<br>Memory:4GB RAM<br>Storage: 250GB HDD |
| Storage Server Pools | Pentium® Dual-Core & Pentium P4 | Processor: Pentium® Dual-Core CPU T4200 @2.00GHz<br>Memory:1GB RAM<br>Storage: 250GB HDD x1<br>Processor: 1.5GHz Pentium P4<br>Memory:512 MB<br>Storage:80GB HDDx20 |

Table 2. Allocated IP address

|  | IP range | Subnet mask |
|---|---|---|
| Cloud Controller (CLC) | 192.168.10.4 | 255.255.255.0 |
| Cluster Controller | 192.168.10.5 | 255.255.255.0 |
| Node Controller (NC) | 192.168.90.0/29 | 255.255.255.0 |
| Virtual Machine Network on Node Controller (NC) | 192.168.40.0/16 | 255.255.255.0 |
| Storage Server Pools | 192.168.50.0/16 | 255.255.255.0 |

## 4. PC CLUSTER BASED STORAGE SERVER DESIGN

PC cluster based storage server architecture is focus of our work. We build terabyte storage server using many of PC cluster nodes.

### 4.1. Physical Topology of CCPS

In this section, we describe how to build physical topology of CCPS. The system framework of CCPS is shown in figure 2.The overall framework of CCPS consists of three layers which are web based application services layer, Hadoop Distributed File System (HDFS) layer and PC cluster layer. The web based application layer provides interface for the users whose can store their own applications such Virtual Machine (VM) images, dataset and multimedia data, etc. The HDFS layer supports the file system for PC cluster layer. HDFS is as a user-level file system in cluster which exploits the native file system on each node to store data. The input data are divided into blocks, typically 64 megabytes and each block is stored as a separate file in the local file system. HDFS is implemented by two services: the NameNode and DataNode. The NameNode is responsible for maintaining the HDFS directory tree, and is a centralized service in the cluster operating on a single node. Clients contact the NameNode in order to perform common file system operations, such as open, close, rename, and delete. The NameNode does not store HDFS data itself, but rather maintains a mapping between HDFS file name, a list of blocks in the file, and the DataNode(s) on which those blocks are stored [8]. The PC cluster layer provides to store large amount of data.

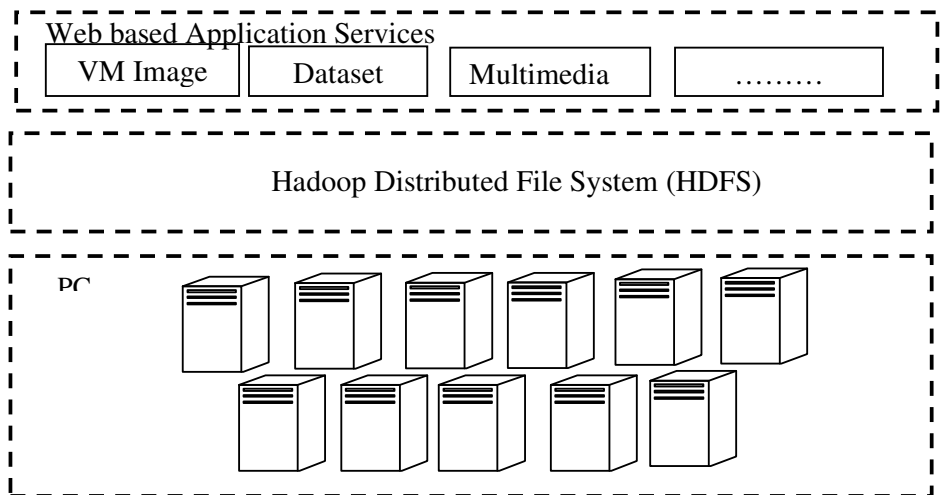

Figure 2: System Framework of CCPS

### 4.2. PC Cluster

Nowadays, many organizations need terabytes storage systems but are very expensive and require higher degree of skills for their operations and maintenance. Usually a desktop PC contain more than 100 GB Hard Disk Drive (HDD), at least 256 MB or greater RAM and 2GHz or higher processor. A typically installation of an operating system and other software installation do not use more than 20 GB of HDD storage. This leaves on the average about 80% of the storage space to be unused. Therefore, I proposed PC cluster based storage server that is inexpensive and easy to maintain.

PC cluster become popular in the 1990 for batching processing of high performance application. A PC cluster is a collection of computer nodes, which is interconnected by a high-speed switching network, all nodes can be used individually or collectively as a cluster.

#### 4.2.1 System overview of PC Cluster based storage server

PC cluster based storage server tries to transfer from the cluster computing to storage server. The storage server uses inexpensive PC components. The large files can be stored by striping the data across multiple nodes. Our system is based on client-server architecture. Each individual machine of a cluster is referred to as a node. In PC cluster consists of one NameNode as server and many DataNodes as clients. CCPS (Cloud based Cluster PC System) uses HDFS (Hadoop Distributed File System) to store data in the collection of the nodes. Each node has its own memory, I/O devices and operating system. The nodes are physically separated and connected via a LAN.

In CCPS consists of a NameNode and DataNodes. NameNode manages the whole PC cluster and maintains the metadata of HDFS that contains the information of blocks, the current locations of blocks, and monitoring the states of all nodes in the cluster. NameNode is recorded any changes to the file system namespace such as opening, closing, renaming files and directories. It also determines the mapping of blocks to DataNodes. The DataNodes store the physical storage of the files. DataNodes also perform block creation, deletion and replication upon instruction from the NameNode. In CCPS, files are divided into blocks that are stored as independent units. Each block is replicated to a small number of separate machines (typically 3) for fault tolerance. In PC cluster, each individual machine of a cluster is referred to as a node. Our system is based on client-server architecture and consists of one server node (NameNode) and many clients nodes(DataNodes) in order to make them work as a single machine. NameNode has two networks cards and one is connect to the front-end infrastructure using gigabyte switch. The PC cluster is the use of multiple computers, typically PCs which was built 1.5 GHz Pentium P4 processors, 80 GB Hard disks and 512 MB RAM and running with Ubuntu OS. These PC are connected by a gigabyte switch that can create illusion of being one machine. The proposed PC cluster based storage server utilizes the existing PC machines in our university without purchasing any extra hardware and software components. Therefore, this storage server is very cost effective architecture.

## 5. ANALYTICAL PERFORMANCE MODEL OF PC CLUSTER BASED STORAGE SERVER

In this section, we propose an analytic performance model for PC cluster based storage server. The storage server typically process many simultaneous jobs (i.e file storage) each of which uses for various shared resources: file access, processor time and network bandwidth. Since one job may use a resource at any time, all other jobs must wait in a queue for their turn at the resource. As jobs receive service at the resource, they are removed from the queue and new jobs arrive and join the queue. Queuing theory is a tool that helps to compute the size of those queues and the time that jobs spend in them. Our model uses M/M/1 queuing network model where M's represent the Markov or memory lass nature of the arrival and service rates and 1 denotes the number of servers attached to the queue, to provide average number of files in the system, average number of service time per file and overall utilization of system. The PC cluster based storage server consists of many PC. The data stored on PC cluster based storage server can access from the academic based private cloud network.

PC cluster based storage server is composed of N independent heterogeneous machines. There is exactly *1* NameNode and the remaining *N-1* machines are DataNodes that storing a total of *B* different blocks $b_1, b_2,..., b_M$ where B>N. NameNode maintain *B* different blocks of metadata M={$M_1, M_2,...,M_n$}, where n is the total number of metadata. Every DataNode $D_i$, there are $B_i$ blocks saved in it and $B_i$= {$b_{i1},b_{i2}, ..., b_{imi}$} is denoted as the set of blocks belonging to $D_i$. We build analytical model for PC cluster based storage server with Poisson arrivals in which I/O requests are issued randomly in time requests have completed.

In model, m application send file request to PC cluster based storage server through IP network. Poisson arrival rates of m request streams are $\lambda$ where the average rate at which new jobs arrive at the queue and the average amount of time that it takes a server to process such jobs is the service time $\mu$ of the server. Let i be the random number of files in the system.

The system can be modelled each state represents the number of files in the system. As the system has an infinite queue and the population is unlimited, the number of states the system can occupy is infinite: state 0 (no files in the system), state 1 (1 file), state 2 (two files), etc. As the queue will never be full and the population size being infinite, arrival rate, $\lambda$, is constant for every state and service rate is also constant for all states. In fact, regardless of the state, we can have only two events (1) a new file arrives. So if the system is in state i, it goes to state i + 1 with rate $\lambda$. (2) a file leaves the system. So if the system is in state i, it goes to state i – 1.

Table 3. Definition of notations

| Symbols | Definition |
|---|---|
| $\lambda$ | Poisson arrival rate |
| $\mu$ | Average service rate |
| i | The number of files in the system |
| $\rho$ | The fraction of time that the server is busy |
| 1-$\rho$ | The fraction of time that the server is idle |
| N | Average number of files in the system |
| W | Expected waiting time in the queue |
| T | Average delay per file |
| $N_Q$ | Average number of files in the queue |

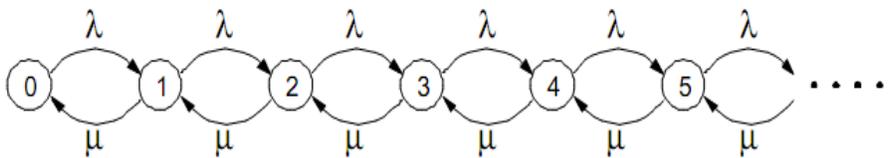

Figure 3. State Transition Diagram

Let $P_i$ = P (System in state i) and the fraction of time that the server is busy should be

$$\rho = \frac{\lambda}{\mu} \quad (1)$$

In particular, $P_0$ is 1-$\rho$ because this is the probability that the queue is empty in steady state. Let N denote the random number of files in the system in the steady state, including the receiving service. The average number of files is defined as in equation 2.

$$N = \sum_{i=0}^{\infty} iP_i = (1- \rho)\sum_{i=0}^{\infty} i\rho^i \quad (2)$$

Consider a typical file entering the system in steady state. Now suppose this newly arriving file sees n files ahead of it. It needs to wait for all of them to complete service and for it own service to complete, before it can depart the system. Hence, the average delay per file (time in queue plus service time) or average response time of the system T is defined as in equation 3.

$$T = \frac{N}{\lambda} = \frac{1}{\mu-\lambda} \qquad (3)$$

The average waiting time in queue W is shown in equation 4.

$$W = \frac{1}{\mu-\lambda} - \frac{1}{\mu} = \frac{\rho}{\mu-\lambda} \qquad (4)$$

The average number of files in the queue $N_Q$ is defined as in equation 5.

$$N_Q = \lambda W = \frac{\rho^2}{1-\rho} \qquad (5)$$

We can also derive performance an estimate of the utilization of the system is showed in equation 6.

$$\text{Utilization} = \sum_{i=0}^{N} \rho \qquad (6)$$

### 5.1 Evaluation and Analysis of the model

This model has two parameters. First, the workload intensity is specified that in this case is the rate at which files arrive (e.g., one file every two seconds). Second, we must specify the service demand is also specified that is the average service requirement of a file. In our validation experiments, assume that evaluate upon different arrival rate of 5, 10, 15, 20, 25, 30 files per second respectively and service rate is 32 second per file. The average response time, utilization and average waiting time of the system with different arrival rate are shown in figure 4, 5 and 6.

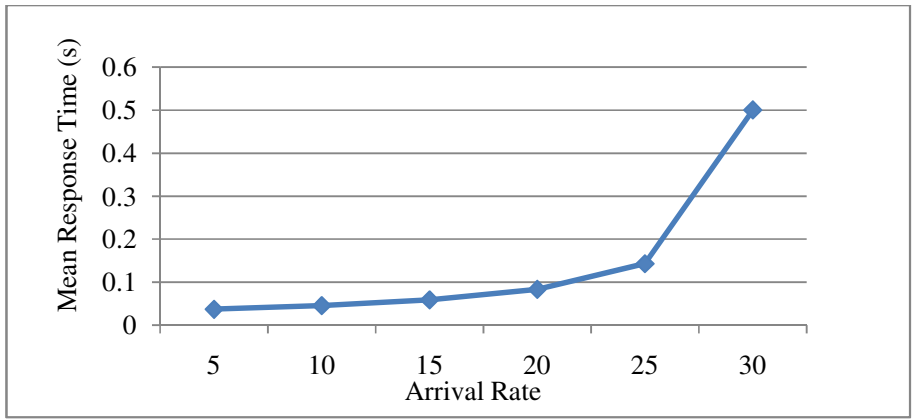

Figure 4. Mean Response Time

By equation 3, the mean response time is 0.03704, 0.04545, 0.05882, 0.08333, 0.14286 and 0.5 respectively. According to the results of figure 4, the result of mean response time depend on the arrival rate, when the arrival rate is increase, the mean response time can be increase.

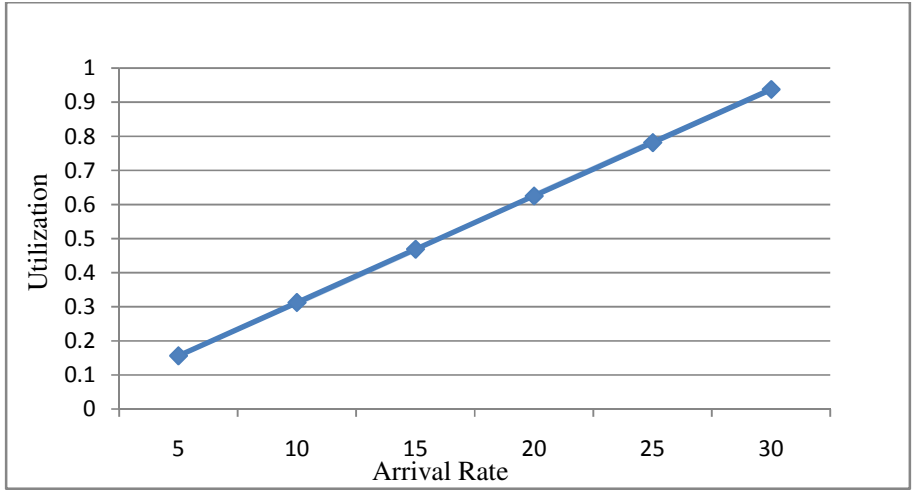

Figure 5. Average Utilization of the Server

By equation 6, the average utilization of the server is 0.15625, 0.3125, 0.46875, 0.625, 0.78125 and 0.9375 respectively. According to the results of figure 5, the average utilization depend on the arrival rate of the system, when the arrival rate is increase, the mean utilization of the server can also be increase.

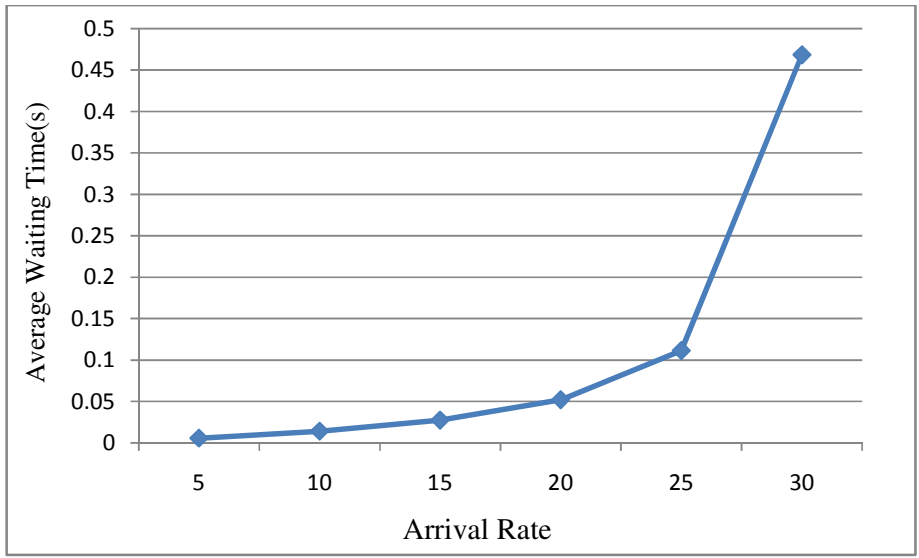

Figure 6. Average Waiting Time

According to the results of figure 6, the average waiting time depending on the arrival rate of the system. Therefore, proposed analytical model is more efficient on the decrease in arrival rate than the increase in arrival rate.

## 6. EXPERIMENTAL ENVIRONMENT

The test platform is built on a cluster with one NameNode and five DataNodes of commodity computer. All nodes are interconnected with 1 Gbps Ethernet network. In each node, Ubuntu server 10.10 with the kernel of version 2.6.28-11-server is installed. Java version is 1.6.0 and Hadoop version is 0.20.2. The size of HDFS blocks is 64 MB and the number of replicas is set to three. The system configuration is shown in Table 4. During the experiments, installation of Ubuntu operating system use 9.8125% and the remaining 90.1875% of the storage space to be used PC cluster based storage server.

Table 4. PC cluster based Storage Server Configuration

| System | Process Configuration | Other Information |
|---|---|---|
| NameNode | Pentium® Dual-Core | Processor: Pentium® Dual-Core CPU T4200 @2.00GHz<br>Memory:1GB RAM<br>Storage: 250GB HDD |
| DataNode1 | Pentium P4 | Processor: 1.5GHz Pentium P4<br>Memory:512 MB<br>Storage:80GB HDD |
| DataNode2 | Pentium P4 | Processor: 1.5GHz Pentium P4<br>Memory:512 MB<br>Storage: 80GB HDD |
| DataNode3 | Pentium P4 | Processor: 1.5GHz Pentium P4<br>Memory:512 MB<br>Storage:80GB HDD |
| DataNode4 | Pentium P4 | Processor: 1.5GHz Pentium P4<br>Memory:512 MB<br>Storage:80GB HDD |
| DateNode5 | Pentium P4 | Processor: 1.5GHz Pentium P4<br>Memory:512 MB<br>Storage:80GB HDD |

In the system configuration used the number of 5 DataNodes and consumed about 10 GB Hard disks storage of each PC for installation of an operating system and other software installation. The available storage capacity of these PCs is combined together, and then can provide 5 x 70 = 350 GB of storage capacity. The storage capacity remains unused and can be utilized if combined to store huge amount of data. If our system configuration used the number of 20 DataNodes, the available storage capacity can provide 20x70=1400 GB. Therefore, storage capacity of the server can be increased depending on the number of PC node in PC cluster. The average storage capacity of PC cluster based storage server is shown in figure 7. Moreover, 100 MB, 100 MB and 1000 MB of data files are stored to the cluster. The results of experiment are shown in figure 8.

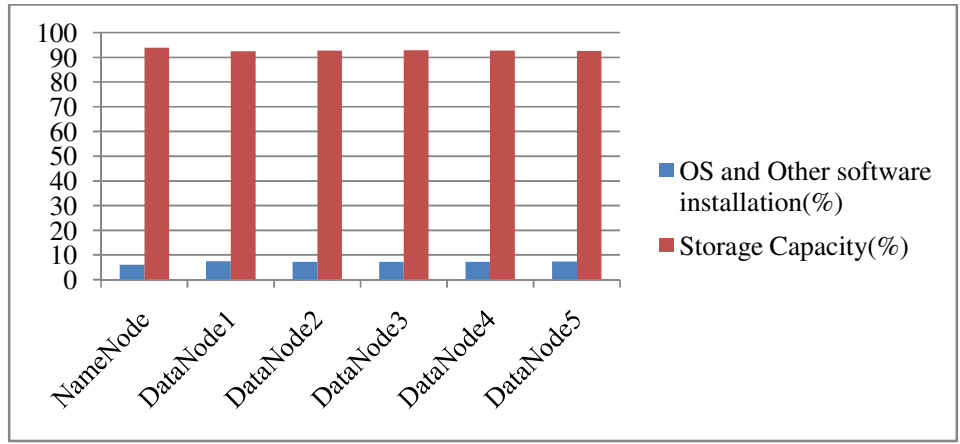

Figure 7. Average Storage Capacity of PC cluster based Storage Server

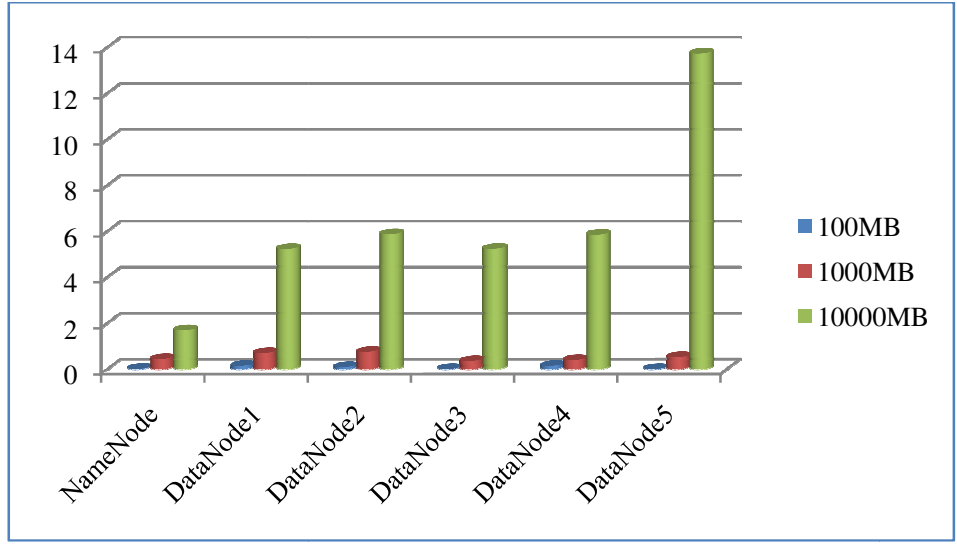

Figure 8. Average Disk Usage of PC cluster based Storage Server

## 7. CONCLUSIONS

In this paper, design and architecture of PC cluster based storage server is configured that utilizes inexpensive computer machines as storage server of the cloud. It can be used to store large amount of data and high performance processing capacity. This paper is proposed an analytical model using M/M/1 queuing network model, which is modelled on intended architecture to provide better response time, utilization of storage as well as pending time when the system is running. As can be seen from experimental results, the storage can be utilized more than 90% of storage space. As future works, the greatest challenges of the system are high scalability and fault tolerance of NameNode.

## ACKNOWLEDGEMENTS

The authors would like to thank everyone, just everyone!

## REFERENCES

[1] Amazon S3.http://aws.amazon.com/s3/

[2] Eucalyptus. http://open.eucalyptus.com

[3] HDFS. http://hadoop.apache.org/core/docs/current/hdfs_design.html

[4] A. Hussam, P. Lonnie, W. Haki , "RACS: A Case for Cloud Storage Diversity", *In Proceedings of the SoCC'10*, 2010.

[5] B. André and E. Sascha, "Inter-node Communication in Peer-to-Peer Storage Clusters", *In Proceedings of 24th IEEE Conference on Mass Storage Systems and Technologies*, 2007.

[6] C.Yong, M.Lionel and Y.Mingyao, "CoStore: A Storage Cluster Architecture Using Network Attached Storage Devices", *In Proceedings of the Ninth International Conference on Parallel and Distributed Systems*, 2002.

[7] G. Garth, "Cloud Storage and Parallel File Systems", *In Proceedings of SNIA Storage Developer Conference*, 2009.

[8] P. Bo, C. Bin and L. Xiaoming, "Implementation Issues of A Cloud Computing Platform", *In Bulletin of the IEEE Computer Society Technical Committee on Data Engineering*, 2009.

[9]   W. Jiyu, Z. Jianlin, L. Zhije, J. Jiehui,   "Recent Advances in Cloud Storage". *In Proceedings of the Third International Symposium on Computer Science and Computational Technology*, 2010, pages 151-154.

[10]  X.Ke, S.Meina, Z.Xiaoqi and S.Junde, "A Cloud Computing Platform Based on P2P". *In Proceedings of IEEE*, 2009.

**Authors**

**Tin Tin Yee.** She received the Bachelor of Computer Science degree in 2005 and Master of Computer Science degree in 2008 from the University of Computer Studies, Yangon, Myanmar. She is currently working toward the Ph.D degree at University of Computer Studies, Yangon, Myanmar. Her research interests in the cloud computing and cloud storage server design.

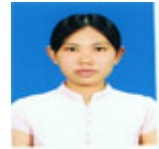

**Thinn Thu Naing.** She obtained her Ph.D degree in Computer Science from University of Computer Studies, Yangon in 2004, and Bachelor of Computer Science degree and Master of Computer Science degree in 1994 and 1997 respectively from University of Computer Studies, Yangon. Currently, she is a Professor in Computer Science at University of Computer Studies, Yangon, Myanmar. Her specialization includes cluster computing, grid computing, cloud computing and distributed computing.

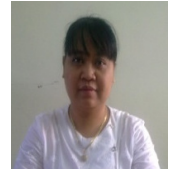